\documentclass[11pt]{article}

\usepackage{amsmath}

\usepackage{amssymb}

\usepackage{epsfig}
\usepackage{amsthm}
\usepackage{latexsym}

\newtheorem {theorem}{Theorem}[section]

\newtheorem{lemma}[theorem]{Lemma}

\newtheorem{proposition}[theorem]{Proposition}

\newtheorem{conjecture}[theorem]{Conjecture}

\newcommand  {\QED}    {\def\qedsymbol{$\blacksquare$}\qed}

\def\tr{{\rm tr}}

\def\D{{\mathcal D}}
\def\R {{\mathcal R}}
\def\C {{\mathcal C}}
\def\T {{\mathcal T}}
\def\G {{\mathcal G}}
\def\E {\mbox{$\mathbb E$}}
\def\Z {\mbox{$\mathbb Z$}}
\def\P{\mbox{$\mathbb P$}}
\def\Q {{\mathcal Q}}

\def\eps{\epsilon}
\def\lam{\lambda}

\newcommand{\m}{{\rm med}}

\newcommand{\path}{{\rm path}}

\begin{document}

\title{How much can evolved characters tell us about the tree that generated them?}
\author{Elchanan Mossel\thanks{Supported by a Miller fellowship in Computer Science and Statistics, U.C. Berkeley.}\\ 
Statistics \\
U.C. Berkeley \\ mossel@stat.berkeley.edu \and Mike Steel 
\thanks{Corresponding author. Thanks to the New Zealand Institute for
  Mathematics and its Applications (Phylogenetic Genomics Programme and Maclauren Fellowship)}\\ 
Biomathematics Research Centre\\
University of Canterbury\\
m.steel@math.canterbury.ac.nz \\
}
\date{\today}

\maketitle

\begin{abstract}
In this chapter we review some recent results that shed light on a fundamental question in 
molecular systematics: how much phylogenetic `signal' can we expect from characters
that have evolved under some Markov process? There are many sides to this question and we begin
by describing some explicit bounds on the probability of correctly reconstructing an ancestral state from 
the states observed at the tips.  We show how this bound sets upper limits on the
probability of tree reconstruction from aligned sequences, and we provide some new extensions that allow site-to-site rate variation or a covarion mechanism. We then explore the relationship between the number of sites required for accurate tree reconstruction and other model parameters - such as the number of species, and substitution probabilities, and we describe a phase transition that occurs when substitution probabilities exceed a critical value. In the remainder of this chapter we turn to models of character evolution where the state space is assumed to be either infinite or very large. These models have some relevance to certain types of genomic data (such as gene order)  and here we again investigate how many characters are required for accurate tree reconstruction.
\end{abstract}

\section{Introduction}
As biologists delve deeper into the evolutionary history of 
life they often find that sequence data provides conflicting or unclear
phylogenetic information. For DNA sequences that have a high site substitution rate
the problem of {\em site saturation} is well known, whereby certain sequences are
essentially random with respect to each other due to the number of substitutions that have occurred during their evolution from a common ancestral sequence.   For
other sorts of data - such a gene order data, where genomes have undergone
much reshuffling  - a similar eventual randomization and loss of information also occurs. 

The phenomenon of randomization, and the rate at which 
it occurs, have been well studied in the probability literature - see for example
Diaconis \cite{dia88}. In this setting it is often useful to regard the stochastic process
as a random walk on a group. For example,
card shuffling, or (unsigned) gene order rearrangement may be viewed as a random walk on
the symmetric group on $n$ elements (i.e. the group consisting of all $n!$ permutations on $n$ elements, under composition) while site substitution in DNA sequences of
length $k$ may be regarded as a random walk on the group $(\Z_2 \times \Z_2)^k$ (since the three types of DNA substitutions -- transitions and the two types of transversions -- together with an identity forms a group under the operation of composition that is isomorphic to group with elements $(0,0), (0,1), (1,0), (1,1)$ under componentwise addition ; a link that was first noted and exploited by Evans and Speed \cite{eva93}).  An alternative setting to a `random walk on a group' is to consider a random walk on 
a finite regular connected graph, and most of the examples we have just mentioned can also be viewed from
this perspective. Either setting - a random walk on a group, or 
a random walk on a graph - is just a special type of ergodic Markov chain, for which the usual questions arise, 
such as what is the limiting distribution, and how fast does the chain approach this
limit?  Often there is an abrupt transition from 
non-random to random in a sense that can be formalized and proved.  For example, with
binary sequences of length $k$ (where $k$ is large) 
under a model of independent site substitution,
this transition occurs when each site has undergone approximately 
$\frac{1}{4}\log_e(n)$ substitutions - beyond this point the derived sequence quickly becomes 
essentially random with respect to the first (for a precise rendition of this
statement see \cite{dia88}, Theorem 3, p. 28).  A similar type of transition for gene order rearrangement under
random inversions was recently derived by \cite{dur03}.

While these questions have been well understood for Markov chains, 
they have been less thoroughly investigated for the more general setting of Markov processes on 
trees. 

The situation here is interesting for the following reason - as the tree gets larger
each leaf tends to become further from the root (and so conveys less information
about the ancestral root state) yet the number of leaves also gets larger. It is, a priori,
not clear whether the gain in information provided by more leaves compensates for the 
losses experienced by each leaf.  This question is also familiar in biology - does 
the sampling of more species provide a strategy for coping with 
site saturation?  As we will see, these  questions are relevant not just for reconstructing ancestral character states, but also for inferring phylogenetic trees.

Evolution processes may be often viewed as Markov processes
on trees. These processes are in turn a special family of Markov random fields
on trees, the study of which is an important branch of 
{\em statistical physics} - see \cite{Georgii} for general background 
and \cite{higuchi:77,EKPS,M:lam2,MosselPeres:03,MaSiWe:03b} for results 
regarding Markov processes on trees. The theory of Markov random fields (and processes) on trees is used to investigate problems such as ancestral reconstruction of states, which is familiar in both biology and physics.
In contrast, the problem of reconstructing the tree topology, which is
well-studied in biology, seems not to have been addressed in the
statistical physics literature.

In this chapter we survey some of the recent advances in the
information-theoretic treatment of Markov processes on trees. We begin
by dealing with Markov processes on a fixed (small) state space - for
example nucleotide sequence data. Here we describe information-theoretic
limits that place bounds on the extent to which ancestral states and
deep divergences can be resolved from sequence data.  We also consider
the question of how much sequence information is required to accurately
reconstruct a tree, a question where there remains an interesting
unresolved issue. We then turn to the analysis of characters on state
spaces that are large or infinite, and which exhibit a somewhat
different (and more tractable) behavior. Along the way we will indicate how such
character data may be relevant to the analysis of genomic data such as
gene order.

\section{Preliminaries} 

In this section we describe some background and notation concerning
phylogenetic trees and Markov processes on trees -
readers familiar with these topics may wish to skim over
this material.

\subsection{Phylogenetic trees}

Throughout this chapter $X$ is a finite set and we will let $n=|X|$.  A
{\em phylogenetic $X$--tree} (or more, briefly, a {\em phylogenetic
  tree}) is a tree $T=(V,E)$ having leaf set $X$, and for which the interior
vertices are unlabelled and of degree at least $3$.  If in addition each
interior vertex has degree exactly $3$ we say that $T$ is {\em
  trivalent}. In evolutionary biology, the set $X$ typically represents the extant species (or sequences)
  while the remaining vertices of the tree represent speciation events (or unknown ancestral sequences). Trivalent trees (also sometimes called `fully-resolved') are regarded as the most informative as they contain no
  `polytomies' (vertices of degree $>3$ that generally represent uncertainty as to the actual order of speciation).

Two phylogenetic $X$--trees $T$ and $T'$ are regarded as equivalent if
the identity map on $X$, regarded as a bijection from the set of leaves
of $T$ to the leaves of $T'$ extends to a graph isomorphism between
the two trees.  Thus, for example, there are precisely three trivalent
(and one non-trivalent) phylogenetic $X$--trees for any set $X$ of size
$4$.  Less formally, two phylogenetic $X$--trees are equivalent if they describe the same 
graphical relationships between the species in $X$, even though the trees might be drawn differently in the plane. 

We are also interested in {\em rooted} phylogenetic $X$--trees. Briefly,
a rooted phylogenetic tree is obtained from a phylogenetic tree by
either distinguishing some interior vertex as a root, or by subdividing
an interior edge and calling the new degree-two vertex a root.  We
denote the root of a rooted phylogenetic tree $T$ by $\rho$, and direct
all edges away from the root. For a rooted phylogenetic tree $T$ we will use throughout this chapter the word 
{\em topology} to denote the associated unrooted phylogenetic tree (obtained from $T$ by
suppressing the root, and if it is of degree $2$ identifying its two
incident edges). A rooted phylogenetic tree is said to be {\em binary}
if each non-leaf vertex has precisely two outgoing arcs. Thus a phylogenetic tree
is binary precisely if its topology is trivalent.  For more background on
the mathematics of phylogenetic trees the reader is referred to \cite{sem03}.

\subsection{Markov processes on trees}

Let $\C$ be the set of character states (such as 
$\C=\{0,1\}, \C = \{A,C,G,T\}$ or $\C = \{20 \mbox{ amino acids}\}$). 
In keeping with biological convention we will
often refer to a site aligned across a set of species $X$ as a {\em character} on $X$; mathematically it is simply a function from $X$ to $\C$. To model the evolution of characters on a rooted phylogenetic tree $T$ by a Markov process we associate to each directed edge $e$ of $T$ 
a matrix $M(e)$ of transition probabilities, and to the root vertex of $T$ we associate a distribution $\pi$ of states (see \cite{erd99} or \cite{ste94} for a more formal description of the model).

Many of the standard models in biology satisfy $M(e) = \exp(t(e)Q)$,
where $Q = (q_{i,j})_{i \in \C, j \in \C}$ is the transition rate matrix
and $t(e)$ represents the `length' of the edge $e$ over which the Markov
process operates. Furthermore, $\pi$ is generally taken to be the
equilibrium distribution that satisfies $\pi Q=0$, so as to induce a
stationary Markov process.

The simplest $2$--state model is the symmetric {\em
  Cavender-Farris-Neyman (CFN) model}
\[
Q = \begin{pmatrix} -1 & 1 \\ 1 & -1 \end{pmatrix},
\]

For this model the probability $p(e)$ of a substitution on any edge $e$ of the tree is
given by
\begin{equation}
\label{subseq}
p(e) = \frac{1}{2}(1-\exp(-2t(e))).
\end{equation}

With $4$ states a slightly more general class of models is
the Tajima and Nei's `equal input' model
\[
Q = \begin{pmatrix} -(a+b+c) & a  & b  & c \\
                        d   & -(b+c+d) & b  & c \\
                        d   & a  & -(a+c+d)  & c \\
                        d   & a  &  b  & -(a+b+d) \end{pmatrix}.
\]
In case $a=b=c=d (=r$, say) this is known as the {\em Jukes-Cantor model}:
\begin{equation} \label{eq:JK}
Q = \begin{pmatrix} -3r & r  & r  & r \\
                        r   & -3r & r  & r \\
                        r   & r  & -3r  & r \\
                        r   & r  &  r  & -3r \end{pmatrix}.
\end{equation} 

Both of these models lead to reversible Markov processes. See \cite{fel04} for various other families of substitution matrices $Q$ appearing in biology.

A further embellishment of most contemporary models of nucleotide substitution is the inclusion of {\em site specific rates}.  That is, one has a distribution $\D$ on some real-valued parameter (the `rate' of evolution of a site) and each site $i$ in the sequence evolves at a rate $\lambda_i$ that is chosen independently from this distribution. 
We refer to the distribution that assigns rate 1 to each site with probability 1 as the {\em degenerate distribution}. 

The substitution process is therefore defined by a transition rate matrix $Q$, a distribution $\D$ of 
site specific rates, a 
rooted phylogenetic tree $T = (V,E,\rho)$, a collection of edge lengths 
$t : E \to \R_{+}$ and a probability distribution $\pi$ on the states at the
root vertex of $T$. 

A {\em configuration} $\sigma : V \to \C$ is a labeling of the vertices of $T$
by $\C$. We will write $\sigma_v$ for the value of $\sigma$ at the 
vertex $v \in V$. The distribution of $\sigma_{\rho}$ is given by
$\pi$. If $u$ is $v$'s parent, then the conditional distribution of $\sigma_v$
given $\sigma_u$ at site $i$ is given by the matrix $M(e) = \exp(\lambda_i t(e) Q)$, where $e =
(u,v)$. We will denote the collection of leaves of the tree $T$ by $\partial T$ and the value of a configuration
$\sigma$ at the leaves by $\sigma_{\partial}$ (which is a character on $X$ - that is, a function from $X$ into the set $\C$).

\section{Information-theoretic bounds: ancestral states and deep divergences}

In this section we describe explicit and easily computable upper bounds
on the information that extant sequences provide concerning (i)
ancestral sequences and (ii) the branching pattern deep inside a tree.
These bounds are in a sense the simplest bounds that can be put on the
reconstruction of ancestral states.

For a leaf $v$, let $\path(v)$ be the set of edges
on the path connecting $v$ to the root $\rho$, 
and let $$t(v) = \sum_{e \in \path(v)} t(e).$$ 
The {\em molecular clock assumption} is that $t(v)$ takes the same value for each $v$; we do not make this
assumption anywhere in this chapter, even though we will refer to sums of $t(e)$ values as (elapsed) `time'.

Let $\pi$ be the prior distribution of the root
character, and let
\begin{equation} \label{eq:defDelta}
\Delta = \sup_{f} \P[f(\sigma_{\partial}) = \sigma_{\rho}],
\end{equation}
be the optimal probability of reconstructing the value of
$\sigma_{\rho}$ given $\sigma_{\partial}$, where the $\sup$ is taken
over all functions. Assuming that the parameters of the model (i.e. $T$,
the $t(e)$ values and the root state distribution $\pi$) are known, it
follows from a classic result (see for example Theorem 17.2 of
\cite{gui}) that an optimal choice of $f$ is the maximum posterior
probability (MAP) estimator - that is, given $\sigma_{\partial}$ one
select the root state(s) $j$ to maximize
$$\P[\sigma_{\partial}|\sigma_{\rho}=j] \cdot \pi[\sigma_{\rho}=j]$$
- a
task that can be carried out by an efficient (polynomial-time in $n$)
dynamic programming approach.

It also follows from standard information-theoretic theory (Theorem 17.3 of \cite{gui}) that the following lower bound on $\Delta$ applies:
\begin{equation} \label{eq:infol}
\Delta \geq 2^{-H(\sigma_{\rho}|\sigma_{\partial})}
\end{equation}
where $H(\sigma_{\rho}|\sigma_{\partial})$ is the 
{\em conditional entropy} of $\sigma_{\rho}$ given $\sigma_{\partial}$
is defined by 
$$H(\sigma_{\rho}|\sigma_{\partial}) = -\sum_{i,
  \sigma}\P[\sigma_{\rho}=i,
\sigma_{\partial}=\sigma]\log_2(\P[\sigma_{\rho}=i|\sigma_{\partial}=\sigma]).$$

Note that in general one cannot expect to recover the root state with
probability close to $1$. Consider for example the Jukes-Cantor model.
Even given the state of the children of the root there is a non-negligible
probability that mutation events occurred along the two edges adjacent to
the root and conditioned on this event the state of the root is
independent from the rest of the character.

As we will see in Section \ref{sec:phase}, there are various
asymptotic results in statistical physics dealing with the limiting behavior
of $H(\sigma_{\rho} | \sigma_{\partial})$ and $\Delta$, but the bounds on
$\Delta$ in most of these results are not explicit. A notable
exception is \cite{EKPS} where a bound on $\Delta$ for the CFN model is
given in terms of ``electrical-resistance'' of an electrical network
defined on the tree.

However our main interest here is in providing explicit \underline{upper} bounds on $\Delta$, which we now describe.  As the rate of substitution increases and/or the temporal separation of the root of the tree from the leaves increases, we would expect it to become increasingly difficult to recover the root state -- a phenomenon well known to biologists as `site saturation'.  However it will be important (particularly for later results) to quantify this rate of decay of information. The following result, which is a slight extension of a result from \cite{mos2}, follows by easy adaptations of coupling arguments appearing earlier in statistical physics, see, e.g., \cite{M:lam2}.  We let
$M_{\D}(x) =\E_{\D}[e^{\lambda x}]$ 
the moment generating function of the site specific rate distribution $\D$. Note that, for the degenerate site specific rate distribution we have $M_{\D}(x) =e^x$. 

\begin{theorem} 
\label{bd:info}
Consider a Markov model on a tree $T$, with transition rate matrix $Q$, edge lengths $t(e)$ (for each edge $e$ of $T$), and site specific rate distribution $\D$.
Let 
\begin{equation} \label{eq:defq}
\begin{array}{ll}
q_j = \min_{i \neq j} q_{i,j}, & q = \sum_j q_j.
\end{array}
\end{equation}
Then the optimal reconstruction probability $\Delta$ for the root state satisfies
\begin{equation} \label{eq:info_bound3}
\Delta \leq \max_i \pi[\sigma_{\rho} = i] + \sum_{v \in \partial T} M_{\D}(-q t(v)).\end{equation}
\end{theorem}

Note that the first term in (\ref{eq:info_bound3}) is precisely the
estimate one would make if one had no knowledge of the character states
at the leaves of $T$. Thus Theorem~\ref{bd:info} says that the improvement
over this `trivial' method decays as the expected exponential of $-qt(v)$. 
Notice also that Theorem~\ref{bd:info}
assumes that $T$ and the values $t(e)$ are all known exactly - if they are not, then the bound on $\Delta$
described applies {\em a fortiori}.

The proof of Theorem~\ref{bd:info} utilizes the method of coupling where one relates one stochastic process to another that is easier to analyze (see e.g. \cite{AF} for background on coupling for Markov chains) . The style of argument employed here has been applied to the study of percolation (see \cite{M:lam2}, and \cite{Pbook,AN} for background). We outline this argument now.  First, we establish the result for the special case of constant
site specific rate, where each site is assigned rate $\lambda$ with probability 1. 
The substitution rate from state $i$ to state $j$ is given by $q_{i,j}$.  Recalling (\ref{eq:defq}), we may define the process equivalently as follows.  Given the current state $i$,
\begin{itemize}
\item[({\bf J1})]
jump to state $j$ with rate $\lambda q_j$;
\item[({\bf J2})]
jump to state $j$ with rate $\lambda(q_{i,j} - q_j)$.
\end{itemize}
The coupling argument relates this process (involving both ({\bf J1}) and ({\bf J2})) to the simpler process involving just ({\bf J1}). 
The crucial point here is that ({\bf J1}) is performed independently of the state $i$.
For edge $e=(u,v)$, let $D(e)$ be the event that a transition of type ({\bf J1}) occurs along the edge $e$. Note that the events $D(e)$ are independent for different edges and that $\P[D(e)^c] = \exp(-q \lambda t(e))$, where $D(e)^c$ denotes the (complimentary) event that $D(e)$ does not occur.
Moreover, conditioned on $D(e)$, 
$\sigma_v$ is independent of $\sigma_{\rho}$. For a leaf $v$, let $D(v)$ be the event that transition of type ({\bf J1})  occurs along an edge $e \in \path(v)$. Then
\[
\P[D(v)^c] = \prod_{e \in \path(v)} \P[D(e)^c] = \prod_{e \in \path(v)} e^{-q\lambda t(e)} = e^{-q \lambda t(v)}.
\]
Finally, let $D$ be the event that $D(v)$ holds for all leaves $v \in \partial T$. Then
\begin{equation} \label{eq:D}
\P[D^c] \leq \sum_{v \in \partial T} \P[D(v)^c] = \sum_{v \in \partial T} e^{-q \lambda t(v)}.
\end{equation}
Note that conditioned on $D$, 
$\sigma_{\partial}$ and $\sigma_{\rho}$ are independent.

To prove the bound on reconstruction (\ref{eq:info_bound3}), 
note that if we are not given
$\sigma_{\partial}$ (or any other information on $\sigma_{\rho}$), then
the best reconstruction function $f$ satisfies $f \equiv j$, where $j$ maximized $\pi[\sigma_{\rho} = i]$ over all $i$,
and this function has success probability $\max_i \pi[\sigma_{\rho} = i]$.
Now let $f$ be any reconstruction procedure and note that, condition 
on the event $D$, $\sigma_{\rho}$ is independent of $\sigma_{\partial}$ and therefore
\begin{eqnarray*}
\P[f(\sigma_{\partial}) = \sigma_{\rho}] &\leq& \P[D^c] + \P[D] \P[f(\sigma_{\partial}) = \sigma_{\rho} | D] \\ &\leq&
\P[D^c] + \P[D] \max_i \pi[\sigma_{\rho} = i] \leq 
\P[D^c] + \max_i \pi[\sigma_{\rho} = i],
\end{eqnarray*}
and so 
\begin{equation}
\label{specialcaseeq}
\P[f(\sigma_{\partial}) = \sigma_{\rho}] \leq  \max_i \pi[\sigma_{\rho} = i] + 
\sum_{v \in \partial T} e^{-q \lambda t(v)}.
\end{equation}

Now, consider the case of a general site specific rate distribution $\D$. Clearly,
$\Delta$ is the expected value (with respect to $\D$) of the conditional probability $\P[f(\sigma_{\partial}) = \sigma_{\rho}|\lambda]$ which we may identify with the left-hand side of
(\ref{specialcaseeq}).  Consequently, 
$$\Delta \leq \E_{\D}[\max_i \pi[\sigma_{\rho} = i]] + \E_{\D}[\sum_{v \in \partial T} e^{-q \lambda t(v)}] =  \max_i \pi[\sigma_{\rho} = i] + 
\sum_{v \in \partial T} M_{\D}(-q \lambda t(v))$$
as required. 
$\QED$

{\bf Example.}
To illustrate Theorem~\ref{bd:info} let us consider the
simplest model on four states, namely the Jukes-Cantor model (\ref{eq:JK}) with a
degenerate site specific rate distribution and a molecular clock. For this model the
equilibrium distribution for states is uniform, so it is natural to take
$\pi[\sigma_{\rho} =i] = \frac{1}{4}$ for all four choices of $i$.  Now
suppose we wish to infer the ancestral state at a vertex in a tree that
was present $t$ years ago, using the states observed now amongst the $n$
extant descendant species.  Theorem~\ref{bd:info}
provides the following bound on $\Delta$:
$$\Delta \leq \frac{1}{4} + ne^{-q t},$$
and we may identify the product $\frac{3}{4}qt$ with the expected number of
substitutions that occur on any path from the root to a leaf. 
For example, if the substitution rate is constant at (say) 1 substitution per million years,
and we have a tree with $n=100$ leaves whose root is at least 10 million
years in the past then $\Delta \leq \frac{1}{4} + 0.0002$ so a character
tells us virtually nothing to help us estimate the state that occurred at
the root.
\qed

Notice that some restriction must be placed on the entries of $Q$ for a
bound such as that given by (\ref{eq:info_bound3}) to be useful. For
example, consider a process with three states, with $\pi[\sigma_{\rho} = i] =
1/3$ for each value of $i$, and with
\[
Q = \begin{pmatrix} -2r & r  & r  \\
                        0   & 0 & 0   \\
                        0  & 0  & 0   \end{pmatrix},
\]
for which $q=0$. 
Then it can be checked that $\Delta \geq 2/3$, however we also have $\max_i
\pi[\sigma_{\rho} = i] = 1/3$ so that for this example, $\Delta$ is always bounded away from $\max_i
\pi[\sigma_{\rho} = i]$.

However Theorem~\ref{bd:info} can be extended to
provide some (exponential-decay) bounds similar to
(\ref{eq:info_bound3}) for certain choices of $Q$ for which $q=0$.  A case in
point is the class of `covarion-type' models (see \cite{gal01},
\cite{pen},\cite{tuf98}) in which each state can either be in an `on'
mode or an `off' mode. A state that is `on' is free to change to other
`on' states, or to turn `off' (at various rates), while a state that is
`off' is only free to turn `on' (at some rate). For two base states and therefore a total of 4 states, namely
$0_{{\rm on}}, 1_{{\rm on}}, 0_{{\rm off}}, 1_{{\rm off}}$ the
corresponding rate matrix $Q$ can be written as:

\begin{equation} \label{eq:on_off_matrix}
Q = \begin{pmatrix} -(r_1+u) & r_1  & u  & 0 \\
                        r_2   & -(r_2+u) & 0  & u \\
                        v   & 0  & -v  & 0 \\
                        0   & v  &  0  & -v \end{pmatrix}.
\end{equation}
and for this matrix it is immediately clear that $q=0$. 

In order to obtain bounds for such models, it is better to apply the coupling 
argument directly to the matrices $M(e)$. Note that, for simplicity, we will also assume 
all the `on' sites undergo substitution at the same rate ($\lambda=1$).
Given any real matrix $A$ let $m_j(A) = \min_i A_{i,j}$ and $m(A) = \sum_j m_j(A)$.
Write $m_j(e)$ for $m_j(M(e))$ and $m(e)$ for $m(M(e))$. 
On the edge $e$, the transition process can be described equivalently as follows: Given the current state $i$,
\begin{itemize}
\item[({\bf J1})]
jump to state $j$ with probability $m_j$;
\item[({\bf J2})]
jump to state $j$ with probability $m_{i,j} - m_j$.
\end{itemize}
Note that, as before, ({\bf J1}) is performed independently of the state $i$. 
Repeating the above argument we thus obtain the following bound on the 
reconstruction probability 
\begin{equation} \label{eq:0_bound}
\Delta \leq \max_i \pi[\sigma_{\rho} = i] + 
\sum_{v \in \partial T} \prod_{e \in \path(v)} (1-m(e)).
\end{equation}
For a given tree and substitution matrices we may apply bound (\ref{eq:0_bound})
directly. However, unlike Theorem~\ref{bd:info}, here it is not enough to know 
for all leaves the total time elapsing from the root. Instead, all the edge
lengths are needed.

More can be said if the process described by 
$Q$ is ergodic (maybe with $0$ entries) so that, for $\epsilon>0$, $\exp(\epsilon Q)$ has all its entries positive. 
Let us assume that the length of all branches is at least $\eps$ and let
$\alpha = \sqrt{1 - m(\exp(\eps Q))}$ and note that $\alpha < 1$.

Note that if $A$ and $B$ 
are two stochastic matrices, then 
$1 - m(A B) \leq (1 - m(A)) (1 - m(B))$. 
Thus, if $t > \eps$, then
\[ 
1 - m(\exp(t Q)) \leq 
1 - m \left(\left( \exp(\eps \lfloor \frac{t}{\eps} \rfloor Q) \right) \right) \leq 
(\alpha^2)^{\lfloor \frac{t}{\eps} \rfloor} \leq 
(\alpha^2)^{\frac{t}{2 \eps}} = \alpha^{t/\eps}.
\]
Substituting this into (\ref{eq:0_bound}) we obtain that 
\begin{equation} \label{eq:bd_0_entries}
\Delta \leq \max_i \pi[\sigma_{\rho} = i] + 
\sum_{v \in \partial T} \alpha^{t(v)/\eps}.
\end{equation}
Note the similarity between this expression and the one in
Theorem~\ref{bd:info}. In particular, in order to apply this bound it
suffices to know for each leaf the total time elapsed from the root.

{\bf Example.}
Consider the case where the rates in (\ref{eq:on_off_matrix})
are given by $r_1=r_2=u=v=1$ per one
million years.
Note that since $Q$ is symmetric the stationary distribution is given
by the uniform distribution.

Assume furthermore that length of all branches is at least $\eps=0.25$.
Using numerical analysis software (e.g. Mathematica) we find that
\[
\exp(0.25 \times Q) = \left( \begin{array}{llll}
0.646645 & 0.156621 & 0.175773 & 0.0209616 \\
0.156621 & 0.646645 & 0.0209616 & 0.175773 \\
0.175773 & 0.0209616 & 0.801456 & 0.00180925 \\
0.0209616 & 0.175773 & 0.00180925 & 0.801456
\end{array} \right).
\]
Therefore $m_1 = m_2 = 0.0209616$ and $m_3 = m_4 = 0.00180925$.
Thus,
$m(\exp(0.25 \times Q)) = 2 \times 0.0209616 +
2 \times 0.00180925 = 0.0419232$ and $\alpha = \sqrt{1 - m} = 0.978814$.

Suppose we now have a tree with $n=100$ leaves and we want to infer the
ancestral state of a state that was present $t$ million years ago.
We thus obtain from (\ref{eq:bd_0_entries}) that
\[
\Delta \leq \max_i \pi[\sigma_{\rho} = i] +  n \alpha^{t/\eps} =
\frac{1}{4} + 100 \alpha^{4 t}.
\]
In particular if $t=100$ million years, the probability of reconstructing
the ancestral state correctly is at most $0.25 + 0.000190498$.
So again, the character reveals essentially no information about the
ancestral
state.
$\QED$

\subsection{Reconstructing deep divergences}

\begin{figure}[htb] 
\input{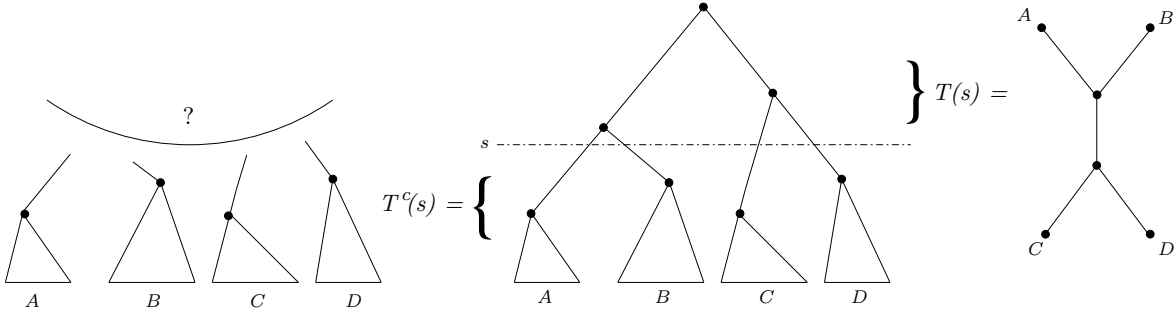} 
\caption{{\em Left:} An example of a deep divergence involving four subtrees. 
{\em Centre and Right:} The tree $T(s)$ and forest $T^c(s)$} 
\label{deepfig}
\end{figure}

 Theorem~\ref{bd:info} allows one to place bounds on the extent to
 which sequences can resolve a divergence event deep inside a phylogeny.
 Consider for example four monophyletic groups of taxa for which we have
 aligned sequences of length $k$. We may wish to determine which of the
 three possible phylogenetic trees connect these four groups, as
 illustrated on the left of Fig.~\ref{deepfig}.

 Clearly, it will only help us in this task if we know the tree
 topologies of each of the four monophyletic groups together with their
 $t(e)$ values.  Each sequence site provides a portion of information
 concerning the `deep' tree structure (i.e. which of the three possible
 phylogenetic trees connect the four subtrees) and
 it is possible to explicitly bound the information
 that the entire sequences provide concerning this divergence. In this
 way one can set explicit lower bounds on the number of sites would be
 needed in order to resolve a deep divergence. One such bound was
 described, for the CFN model, in \cite{sob03}. Here we describe a more
 general approach from \cite{mos2} that applies to a wider range of
 models and settings.
 
 Let $T(s)$ denote the  topology of tree $T$ up to time $s$
 from the root, and let $T^c(s)$ denote the forest consisting of the 
 subtrees from time $s$ to the present (including the associated edge lengths). In other words, $T(s)$ describes
 all divergences up to time $s$, while $T^c(s)$ describes all divergences (and their relative separations)
 from time $s$, as illustrated on the center and right of Fig.~\ref{deepfig}.

Consider the problem of reconstructing $T(s)$ (given $T^c(s)$) from a sequence of characters
that are generated by a common Markov process on $T$, where
the prior distribution on $T(s)$ is given by a measure $\mu$.
The prior $\mu$ is on $T(s)$ with its edge lengths. However, for a tree topology $T$, we will 
write $\mu[T(s) = T]$ for the prior probability that the topology of $T(s)$ is given by $T$. 

Note that in the following result (Theorem~\ref{boundingthm}) 
we do not need to assume independence between sites that evolve according to this process
on $T$.

Let us denote a sequence of $k$ identically generated configurations by $\sigma^1,\ldots,\sigma^k$. 
We will also denote the values of the configuration
$\sigma^i$ at the leaves by $\sigma^i_{\partial}$.
Similarly, we denote by $\sigma^i_{\rho}$ the value of the configuration
$\sigma^i$ at the root $\rho$. Suppose furthermore, that the characters evolve as in 
Theorem~\ref{bd:info} with substitution matrix $Q$;
and we have a site specific rate distribution $\D$. 
Let $\Delta^T(s)$ be the probability of reconstructing, given
$T^c(s)$ (with its associated $t(e)$ values) the tree
topology up to time $s$,
\begin{equation} \label{eq:defDelta2}
\Delta^T(s) = 
\sup_{f} \P[f((\sigma_{\partial}^j)_{j=1}^k) =  T(s) |T^c(s)].
\end{equation} 
The $\sup$ is taken over all functions, and as before, the optimal
choice of  $f$ is the maximum posterior probability (MAP) estimator,
which given $(\sigma_{\partial}^j)_{j=1}^k$ selects a tree $T'$ to
maximize 
\[
\int 1_{\{T(s) = T'\}} \P[(\sigma_{\partial}^j)_{j=1}^k|T(s),T^c(s)] d \mu(T(s)),
\]
(where $1_{\{T(s) = T'\}}$ is $1$ or $0$ depending on whether
 the topology of $T(s)$ is  $T'$ or not). 
Clearly the probability of reconstructing $T$
from $(\sigma_{\partial}^j)_{j=1}^k$ is less or equal to $\Delta^T(s)$;
this latter quantity, which is the probability of correctly determining
the `deep' part of the tree, can be bounded as follows.

\begin{theorem}
Suppose that $k$ sites  evolve under a Markov process with a site specific
rate distribution $\D$. Then, for any $s>0$ we have:
\label{boundingthm}
\begin{equation} \label{eq:rec_top}
\Delta^T(s) \leq \max_T \mu[T(s) = T] + k \sum_{v \in \partial T}
M_{\D}(-q (t(v)-s)),
\end{equation}
where $q$ is given by (\ref{eq:defq}).
\end{theorem}

{\em Outline of the proof.}
The argument follows similar lines to the proof of
Theorem~\ref{bd:info}.  For character $i$ we say that event $D_i$ occurs
if, for all $v \in \partial T$ there exists a time $t \geq s$ at which a
transition of type ({\bf J1}) occurs at least once on the path
connecting $v$ to the
root of the component of $T^c(s)$ that contains $v$.  By the proof of
Theorem~\ref{bd:info} it follows that $$\P[D_i^c|\lambda_i] \leq \sum_{v
  \in \partial T} e^{-\lambda_iq(t(v)-s)},$$ where $\lambda_i$ is the
rate (chosen from $\D$) that site $i$ evolves at. Consequently,
$$\P[D_i^c] \leq \sum_{v
  \in \partial T} M_{\D}(-q(t(v)-s)),$$ and so, by the Bonferroni inequality,
$$\P[(\cap_{i=1}^kD_i)^c] \leq k \sum_{v
  \in \partial T} M_{\D}(-q(t(v)-s)).$$
Now, conditional on $\cap_{i=1}^kD_i$, the two random variables
  $(\sigma_s^i)_{i=1}^k$ and $(\sigma_{\partial}^i)_{i=1}^k$ are
  independent, and therefore, $T(s)$ and $(\sigma_{\partial}^i)_{i=1}^k$
  are independent.  As in Theorem~\ref{bd:info} we conclude that
$$\Delta^T(s) \leq \P[(\cap_{i=1}^kD_i)^c] + \P[(\cap_{i=1}^kD_i)]\max_T
\mu[T(s)=T] \leq k\sum_{v \in \partial T} M_{\D}(-q(t(v)-s)) + \max_T
\mu[T(s)=T],$$
as required. 
\qed

{\bf Example.} To illustrate Theorem~\ref{boundingthm} let us consider
again the Jukes-Cantor model (\ref{eq:JK}), with a degenerate site specific rate
distribution and molecular clock.  Suppose we have four monophyletic groups of taxa - each
with 100 extant species, and with a well-specified tree with edge lengths
- and we wish to determine which of the three possible trees (choices
for $T(s)$) describes how the trees are joined ancestrally (as in
Fig.~\ref{deepfig}).  In the absence of any prior information it is
natural to take $\mu[T(s) = T'] = \frac{1}{3}$ for each of the 
three possible trivalent trees $T'$.  Suppose it is believed that all four
lineages existed as far back as (at least) 1 billion years ago, and
taking (for example) a site substitution rate ($3r$) of one substitution
per fifty million years, we have  for any leaf $v$ that $qt(v) = 4rt(v) = \frac{4}{3} \cdot (3r) t(v) = \frac{4}{3} \cdot 20$. 
Theorem~\ref{boundingthm} then gives $\Delta^T(s)
\leq \frac{1}{3} + 100ke^{-26.7}$ which implies that at least 700 million sites (!)
would be required in order to have any hope of estimating the ancestral
divergence with probability more than about 0.5. This is perhaps not too surprising given that
the expected number of substitutions per site along the path from the root to any leaf is $20$. \qed

{\bf Remarks}

\begin{itemize}
\item[{\bf (1)}]
As noted above, Theorem~\ref{boundingthm} applies even when the sequence sites are not independent. 
It is possible to extend this theorem further to allow the sites to evolve
according to different Markov processes. 

\item[{\bf (2)}]
In order to get a feeling for the asymptotic behavior of (\ref{eq:rec_top}),
fix $s$ and assume that the tree has $n = e^{\beta t}$ 
leaves, all at time $t$. Here we take the asymptotics where 
$t \to \infty$ (and therefore $n \to \infty$), while $s,q$ and $\beta$ are
all constants. Also we assume a degenerate site specific rate distribution.
Then
\[
\sum_{v \in \partial T} e^{-q (t(v)-s)} = \exp(s q) \exp(-t (q - \beta)).
\]
Therefore if $q > \beta$, then by (\ref{eq:rec_top}) if we want to reconstruct the topology up to time $s$ with high
probability, i.e., $\Delta^T(s) \geq \max_T \mu[T(s) = T] +
\delta$, where $\delta>0$ then we need that
\[
k \geq \delta \exp(-sq) \exp(t (q - \beta)) = \delta \exp(-sq) n^{q/\beta - 1}.
\]
So the number of characters required grows polynomially with $n$.
\end{itemize}
\qed

\subsection{Connection with information theory}

Similar bounds to the ones we have described so far can also be stated
and derived using classical information theory. First we briefly recall
the concept of mutual information. For random variables $X$ and $Y$
the {\em mutual information} between $X$ and $Y$ is defined by
 $$I(X;Y) (= I(Y;X)) := \sum_{x,y}\P[X=x,Y=y] \log_2 \left(
  \frac{\P[X=x,Y=y]}{\P[X=x]\P[Y=y]}\right).$$

Formally, $I(X;Y)$ is the Kullback-Leibler separation of the joint
distribution of $X,Y$ and the product distribution of $X$ and $Y$.
Consequently, $I(X;Y) \geq 0$ with equality if and only if $X$ and $Y$
are independent. Informally $I(X;Y)$ measures the amount of information
that $Y$ carries about $X$ (or conversely that $X$ carries about $Y$).
When $I(X;Y)$ is small then the best method for inferring $Y$ from $X$
does little better than the best method that simply ignores $X$ - a
precise formalization of this claim is Fano's inequality (see \cite{CT}
for more details).

The quantity $I$ has some generic properties that make it  useful
for analyzing the information loss of Markov processes. For example,
suppose that $X,Y$ and $Z$ be random variables such that $X$ and $Z$ are
  independent given $Y$. Then $I(X;Z) \leq \min\{I(X;Y),I(Y;Z)\}$ (the
  `data processing lemma') and $I((X,Z);Y) \leq I(X;Y) + I(Z;Y)$ (the
  `subadditivity property'). By exploiting these properties one can
  derive information-theoretic analogues of 
Theorems~\ref{bd:info} and \ref{boundingthm}
 which we will now briefly describe.
For convenience we will deal just with the degenerate site distribution
in both cases. In the setting of Theorem~\ref{bd:info} it can be shown that
$$I(\sigma_{\partial}; \sigma_{\rho}) \leq \log_2|\C|\sum_{v \in
  \partial T} e^{-qt(v)}.$$
Similarly, in the setting of Theorem~\ref{boundingthm} it can be shown
that 
$$I(T(s); (\sigma_{\partial}^j)_{j=1}^k|T^c(s)) \leq k \sum_{v \in
  \partial T} e^{-q(t(v)-s)}.$$
For
further details, and applications of these results, see \cite{mos2}.

The results described in this section may give the impression that phylogenetic information decays in a smooth fashion according to an interplay of time, substitution rate, and numbers of leaves in the tree.  However as we explain in the next section there are underlying transitions in this behavior.

\section{Phase transitions in ancestral state and tree reconstruction} 
\label{sec:phase}

There is an interesting change (`phase transition') in the behavior of Markov models of
character evolution on trees as the probability of substitution on edges of the trees passes a certain
critical value. This has been well studied in statistical
physics and in information theory, in the context of broadcasting on
trees. But it is also relevant to biology - particularly in attempting
to recover information (ancestral states, branching order) deep within a
tree, from observing the character states at the leaves.

The transition is most easily explained, and has been most studied for
the case of the $2$--state symmetric process (the CFN model described
above).

\begin{figure}[htb] 
\begin{center}
\input{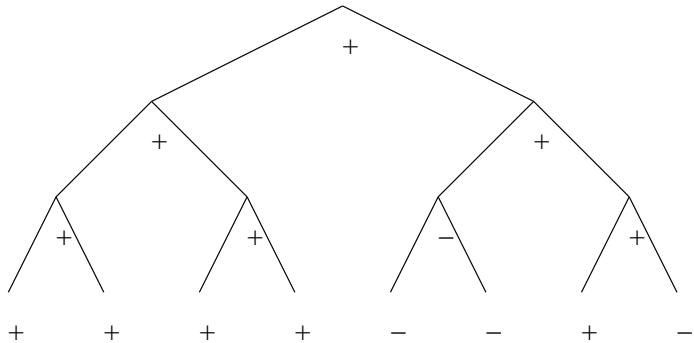} 
\caption{A character of the CFN process on a binary phylogenetic tree on $8=2^3$ leaves at distance $3$ from the root}  \label{rootstate}
\end{center}
\end{figure}

To illustrate this transition between what is called the `ordered' and
`unordered' phases of a Markov process on a tree, suppose we have a
rooted binary phylogenetic tree $T$ that has $n=2^m$ leaves that are at
distance $m$ from the root vertex, as indicated in Fig.~\ref{rootstate}.

Under the CFN model (and with a degenerate site specific rate
distribution) let $$\theta(e) := \det(M(e)) = \det(\exp(t(e)Q)),$$
where, for any square matrix $M$, $\det(M)$ denotes the determinant of $M$ (the product of the eigenvalues of $M$).
A classic identity in linear algebra, Jacobi's identity, states $\det\exp(M) = \exp(\tr(M))$ where $\tr(M)$ is 
the trace of $M$ (the sum of the diagonal entries of $M$, which also equals the sum of the eigenvalues of $M$).  Thus,
$$\theta(e)= \exp(t(e)\tr(Q)) = e^{-2t(e)}.$$
By (\ref{subseq}) we
have $\theta(e) = 1-2p(e)$.  Now suppose that each edge of $T$ has the
same $t(e)$ value, say $t$, and thereby the same $\theta(e)$ value,
namely $\theta = \exp(-2t)$.

Let us further suppose that the distribution $\pi$ of states at the root is uniform 
(i.e. a fair coin toss) and that we wish to use the states $\sigma_\partial = (\sigma_{\partial}^i)$ at the leaves of $T$ to estimate
the state $\sigma_\rho$ at the root. This gives rise to an interesting contest as $m$ (the height of the tree) increases - firstly, each leaf is becoming increasingly far from the root, and so the information that it carries
about the ancestral root state decays to $0$ with increasing $m$.  On the other hand, the number of leaves 
grows (exponentially) with $m$, and so although each leaf carries less information, it might be hoped that together they compensate for their individual losses.  Which factor wins out depends critically on the value of $\theta$. 
Evans {\em et al.} \cite{EKPS} established that, for $2\theta^2 <1$ the mutual information
$I(\sigma_\partial, \sigma_\rho)$ converges to 0, as $m$ tends to infinity (this result was first proven independently by \cite{BRZ} in  a different formulation).  Thus, eventually (as the root becomes increasingly `deep' in the tree) it becomes impossible to estimate the root state with any better success than a blind guess, when $\theta$ lies in this region.  On the other hand, when $2\theta^2 >1$ then $I(\sigma_\partial, \sigma_\rho)$  is bounded away from 0, so that information about the root `survives' to the leaves, no matter how large the tree grows. In this case
maximum likelihood estimation or majority rule estimation (i.e. select the root state that corresponds to 
the majority state at the leaves) suffices to recover some information right up to (but not including \cite{pp}) the critical value $2\theta^2=1$. 

Notice that this critical value translates to a common $t(e)$ value of 
$t = \frac{1}{4}\log(2)$ and thereby to a common $p(e)$ value of $p = \frac{1}{2}(1- \frac{1}{\sqrt{2}}).$ 

Curiously, the maximum parsimony approach for ancestral state reconstruction 
(i.e. select the root state that requires the 
fewest transitions to account for the leaf states) recovers information under the CFN model for values of $p$ only up to $\frac{1}{8}$ \cite{cha95}.

The situation for $r$-states models and for non-symmetric $2$-state
processes is more subtle. There is not any general criteria for deciding
when the mutual
information $I(\sigma_{\partial};\sigma_{\rho})$ is
 converging to $0$ and when is it bounded away from $0$. In fact,
such criteria do not even exist
for symmetric processes on more than $2$ states or for general processes
on $2$ states. In the general setting, there are
various conditions which imply that the
mutual information either converges to
$0$ or is bounded away from $0$. However, these conditions are not sharp.
We describe an example of both types of  conditions now.

Suppose that $M(e) = M$ for all $e$. Since
$M$ is a stochastic matrix, $1$ is an eigenvalue of $M$.
Let $\{1 = \lam_1,\ldots,\lam_r\}$ denote the set of eigenvalues of $M$
and let $\theta = \max\{ |\lam_2|,\ldots,|\lam_r|\}$
(note that for the CFN model, this is consistent with the previous
definition of $\theta$).
A ``spectral criterion'' (\cite{KestenStigum:66,MosselPeres:03}) implies
that for any $M$ if
$2 \theta^2 > 1$ then $I(\sigma_{\partial};\sigma_{\rho})$ is bounded away
from zero for all trees. This result is not tight in general
(see \cite{M:lam2,MosselPeres:03,JansonMossel:04}).

In order to illustrate the spectral criterion
consider the Jukes-Cantor model (\ref{eq:JK}).
Note that the eigenvalues of $Q$ are $0$ (with multiplicity $1$) and
$-4r$ (with multiplicity $3$). Thus if $M = \exp(t Q)$,
then the eigenvalues of $M$ are
$1$ and $e^{-4 r t}$. Therefore, if the stochastic matrix $M = \exp (t Q)$
satisfies
\begin{equation} \label{eq:JK_cond}
A := 2 e^{-8 r t} > 1
\end{equation}
then by the spectral criterion
$I(\sigma_{\partial};\sigma_{\rho})$ is bounded away from zero for all
trees.
This should be compared to Theorem \ref{boundingthm} which implies that if
\begin{equation} \label{eq:JK_cond2}
B := 2 e^{-4 r t} < 1
\end{equation}
and the tree is of depth at least $d$, then the probability of
reconstructing
the ancestral states is bounded by $1/4 + (2 e^{-4 r t})^d$. The
expression
$1/4 + (2 e^{-4 r t})^d$ converges to $1/4$ when $d \to \infty$.
Thus, condition (\ref{eq:JK_cond}) implies that then
$I(\sigma_{\partial};\sigma_{\rho})$ is bounded away from zero for all
trees, while condition (\ref{eq:JK_cond2}) implies that
$I(\sigma_{\partial};\sigma_{\rho})$ converges to $0$
as $d \to \infty$.

In the other direction,
various conditions are derived in
\cite{M:lam2,MosselPeres:03,MaSiWe:03b,Martin:03}
that imply that $I(\sigma_{\rho};\sigma_{\partial})$ converges to $0$ for
various processes.
The simplest of these conditions is given in \cite{M:lam2} -
this condition is closely related to the one given in Theorem
\ref{boundingthm}.
The results in \cite{MosselPeres:03,MaSiWe:03b,Martin:03} give sharper
bounds for symmetric processes on more than $2$ states and for general
$2$--state processes.

Let us illustrate how Proposition 4.2 of \cite{MosselPeres:03}
translates to the ``Jukes-Cantor'' setting. This proposition
specialized for the Jukes-Cantor model asserts that if
\begin{equation} \label{eq:JK_cond3}
C := \frac{2e^{-8rt}}{\frac{1}{2} + \frac{e^{-4rt}}{2}} < 1
\end{equation}
then $I(\sigma_{\partial};\sigma_{\rho})$ converges to $0$
as $d \to \infty$. Simple algebra shows that $A \leq C \leq B$.
Thus (\ref{eq:JK_cond3}) gives a weaker
condition than \ref{eq:JK_cond2}) (and therefore a stronger result)
implying that
$I(\sigma_{\partial};\sigma_{\rho})$ converges to $0$.

\subsection{The logarithmic conjecture}
\label{logsec}

Suppose we generate $k$ characters independently and according to 
the CFN model (with degenerate site specific rate distribution), and ask how large $k$ should be in order that, with probability at least $1- \epsilon$  we can correctly recover from these characters the topology of the underlying phylogenetic tree.  Let $k_{\min}(\eps)$ be the smallest value of $k$ that achieves this last property. Clearly $k_{\min}(\eps)$ depends on features of the generating tree, in particular
the number $n$ of leaves, and the assignment of $t(e)$ values to the edges of this tree (it also depends on $\epsilon$, however we will regard this as a fixed small number). Any dramatic `shortening' of an interior edge, or `lengthening' of an exterior edge (i.e. making the $t(e)$ value small or large, respectively) will cause $k_{\min}(\eps)$ to diverge and so we will assume that each binary phylogenetic tree has all its $t(e)$ values in some fixed interval $[l_n, u_n]$ which may depend on $n$.  The questions of interest are then to determine
the dependence of $k_{\min}(\eps)$ on $n$ and the values $(l_n, u_n)$.  Essentially this question provides another formalization of the question `how much phylogenetic information is contained in characters that evolve according
to a simple Markov model.'  The authors of 
\cite{erd99} showed that, 
\begin{equation}
\label{keq}
k_{\min}(\eps) \leq c'_{\epsilon}  \cdot \frac{\log(n)}{l_n^2} \cdot \exp(u_n\delta_n(T))
\end{equation}
where $c'_{\epsilon}$ is a constant (dependent only on $\epsilon$) and $\delta_n(T)$ is a function (only) of the phylogenetic tree $T$ and that grows slowly with $n$.  Specifically,
$\delta_n(T)$ is at most a constant times $\log(n)$, but is typically (i.e. on average) $O(\log(\log(n))$. It is a measure of how many edges of the tree separate the `deepest' vertex from its nearest leaf.

Thus if we were to regard $l_n$ and $u_n$ as constants (independent of $n$) then $k_{\min}(\eps)$ is at worst polynomial in $n$, and more typically a power of $\log(n)$ (improving an alternative bound described in \cite{far96}). We have not mentioned the tree reconstruction method used to establish (\ref{keq}); it is a polynomial time (in $n=|X|$) algorithm, and chosen more for tractability of analysis than for any supposed superior 
performance; a comparable analysis for maximum likelihood seems more difficult \cite{ste2}.

An obvious question arises: is the bound on $k_{\min}(\eps)$ described by (\ref{keq}) and the consequent relationship between $k_{\min}(\eps)$ and $n$ (for $l_n, u_n$ fixed) optimal?  Certainly $k_{\min}(\eps)$ must grow at least as fast as (a constant times) $\log(n)$, by elementary counting arguments.  This applies under any model of sequence evolution on a bounded state space and any tree reconstruction method \cite{sze}.
The essence of this argument is the following: there are $\frac{(2n-4)!}{(n-2)!2^{n-2}}$ trivalent phylogenetic $X$--trees
and $r^{nk}$ collections consisting of $n$ aligned sequences  of length $k$ on an $r$--letter alphabet, and so if $k = o(\log(n))$ then
for sufficiently large $n$ there exist more trivalent phylogenetic $X$--trees than
$r$--letter sequences of length $k$.

Also an inverse square dependence of $k_{\min}(\eps)$ on $l_n$ is necessary, even when $n=4$, as shown in 
\cite{ste2}. However there is reason to believe that (\ref{keq}) is not optimal, provided that $u_n$ is less than the critical transition value (viz. $\frac{1}{4}\log_e(2)$) between the ordered and unordered states, discussed above. This has led to the following conjecture, which promises a remarkable strengthening of (\ref{keq}) under a further restriction.

\begin{conjecture}
\label{conj}
Consider the CFN model for binary characters, and suppose that 
$u_n \leq u<  \frac{1}{4}\log_e(2)$. Then
$$k_{\min}(\eps) \leq c_{\epsilon, u} \cdot \frac{\log(n)}{l_n^2},$$
where $c_{\epsilon, u}$ is a constant that depends only on $\epsilon$ and $u$. 
\end{conjecture}

Conjecture~\ref{conj} is clearly true for trees for which $\delta_n(T)$ is bounded -- these are trees for which no vertex is very far from a leaf (for example, the class of `caterpillar trees' which are the trivalent phylogenetic trees tfor which every interior vertex is adjacent to a leaf). However for trees that have `deep' vertices such as the complete balanced binary phylogenetic tree that has all its $n=2^m$ leaves at distance $m$ from a fixed central edge, bound (\ref{keq}) is polynomial in $n$. Yet precisely in this
`worst case' setting the bound promised by Conjecture~\ref{conj} holds - this was recently established in \cite{mos}, using an entirely different approach from \cite{erd99}.   The paper \cite{mos} also showed that the restriction on $u_n$ is necessary for Conjecture~\ref{conj} to hold, for when $u_n$ is allowed to take larger values, polynomial dependence of $k_{\min}(\eps)$ on $n$ can result.

Conjecture~\ref{conj} has been extended to a much more general conjecture in \cite{mos} 
concerning the transition from logarithmic to
polynomial dependence of $k_{\min}(\eps)$ on $n$ for a range of 
Markov models at the corresponding transition from the ordered to unordered phase of the process. 

\subsection{Reconstructing forests}

Given that it may be difficult to reconstruct deep parts of a tree (for example, in the region where 
Conjecture{conj} does not apply) a more modest task may be to 
to reconstruct most of the tree that is not `deep'. A natural question then is whether this can be 
achieved using a small (logarithmic in $n$) number of sites).

Note that for the rooted binary tree on $2^m$ leaves, where all the leaves
are at distance $m$ from the root, only an $O(2^{-s})$ fraction of the
vertices is at distance $s$ or more from the set of leaves.
This is true in general for binary trees - only a $O(2^{-s})$ of the
vertices
are at distance $s$ or more from the set of leaves. In other words, for
all
binary trees the ``deep'' part consists of exponentially small fraction
of the tree. Therefore reconstructing the part which is not ``deep''
still contains a lot of information on (recent) divergences.

It turns out that the answer to the above question is positive.
In \cite{m:forest} it is shown that a logarithmic number of characters
suffices to reconstruct a forest containing most edges of the tree.
Moreover, \cite{m:forest} gives a formula relating the ``depth'' of the
forest that can be recovered from a given number of characters.

\section{Processes on an unbounded state space: The random cluster model}

For the remainder of this chapter we will investigate the phylogenetic information that is provided by models which have a large state space.  In this section we deal with a slightly idealized `random cluster' model, in which the underlying state space might be regarded as being infinite - it has the property that any substitution always gives rise to a new state. We will see that this simple model is quite tractable and leads to a result (Theorem~\ref{mainthm}) that is much cleaner than anything that has been established yet for even the CFN model. We will apply this result in the final section of this chapter to investigate a class of models on large but finite state spaces.

For the type of Markov model on a small state space that we have dealt with so far the subsets of the
vertices of a phylogenetic tree $T$ that are assigned particular states
do not generally form connected subtrees of $T$ (in biological
terminology this is because of `homoplasy' - the evolution of the same
state more than once in the tree, either due to reversals or convergent evolution).

However increasingly there is interest in genomic characters such as
gene order where the underlying state space may be very large
(\cite{gal02}, \cite{mor1}, \cite{mor2}, \cite{rok}). For example, the
order of $k$ genes in a signed circular genome can take any of
$2^k(k-1)!$ values. In these models whenever there is a change of state
- for example a re-shuffling of genes by a random inversion (of a
consecutive subsequence of genes) - it is likely that the resulting
state (gene arrangement) is a unique evolutionary event, arising for the
first time in the evolution of the genes under study.  Indeed Markov
models for genome rearrangement such as the (generalized) Nadeau-Taylor
model \cite{mor1}, \cite{nad}  confer a high probability that any given
character generated is homoplasy-free on the underlying tree, provided
the number of genes is sufficiently large relative to $n=|X|$
(\cite{sem02}). Here the phrase 
`homoplasy-free' refers to the condition that a character has parsimony
score equal to the number of states it takes at the leaves minus 1;
this condition has a natural interpretation in biology, since it 
is equivalent to requiring that the character could
have evolved on the tree without reversals or convergent evolution (for
 details of that connection see \cite{ sem02, sem03}). In this setting a  `random cluster' model which we will describe here is the
appropriate (limiting case) model, and may be viewed as the phylogenetic
analogue of what is known in population genetics as the `infinite
alleles model' of Kimura and Crow \cite{kim64}.

Thus for this section we consider the size of the state space to be infinite (or at least very large, and perhaps variable with $n$).  Some of the arguments described above are no longer valid
in this setting. For example, the simple argument in Section~\ref{logsec} that showed that $k_{\min}(\epsilon)$ must grow at least 
as fast as the function $\log(n)$ does not apply when the size of
state space is infinite, or finite but variable with $n=|X|$.  Indeed it
has recently been shown that for any trivalent phylogenetic $X$--tree
$T$ there is an associated set of just {\em four} characters for which
$T$ is the only phylogenetic $X$--tree on which each character in 
that collection has a homoplasy-free evolution (see \cite{sem02}, \cite{hub}). 
Thus it is reasonable to ask whether $O(1)$ characters
might suffice to reconstruct $T$ under a simple random model. We will see that
the answer to this question is `no', but clearly we need a different type of argument.

\begin{figure}[htb] 
\begin{center} 
\input{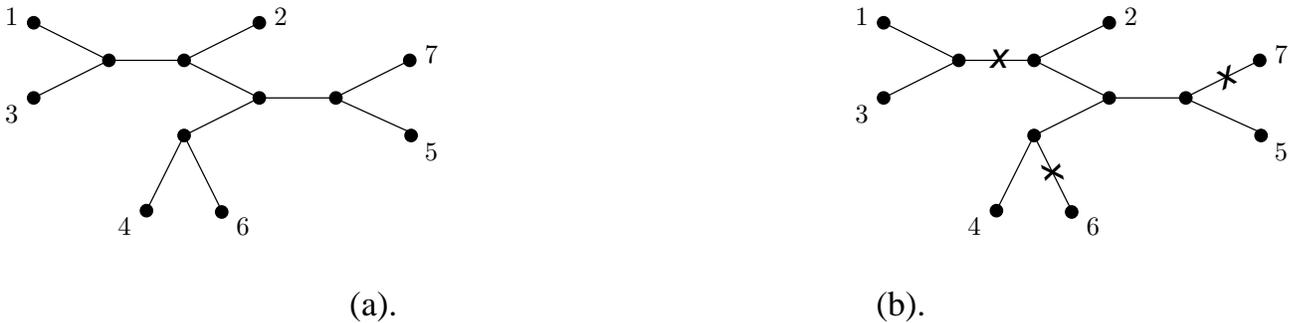} 
\caption{(a) A 
trivalent phylogenetic $X$--tree $\T$ for $X=\{1,2,\ldots, 7\}$;
    (b) For the random cluster model, cutting the edges of $\T$ that
    are marked by a cross  induces the character $\chi$ on $X$ given by
     $\chi=\{\{1,3\}, \{2,4,5\}, \{6\}, \{7\}\}$.} 
    \end{center}
\label{fig3} 
\end{figure}

Consider the following random process on a phylogenetic tree $T$. 
For each edge $e$ let us independently either cut this edge - with 
probability $p(e)$ - or leave it intact. The resulting disconnected graph 
(forest) $G$ partitions the vertex set $V(T)$ of  $T$ into non-empty sets 
according to the equivalence relation that $u \sim v$ if $v$ and $v$
are in the same component of $G$.  This model thus generates random 
partitions of $V(T)$, and thereby of $X$ by connectivity, 
and we will denote these partitions of $V(T)$ and $X$ using the symbols
$\overline{\chi}$ and $\chi$, respectively.  Fig.~\ref{fig3}(b) illustrates this process. 

For an element $x\in X$ we will let
$\chi(x)$ denote the equivalence class containing $x$. 
We call the resulting probability distribution on partitions of $X$ the
{\em random cluster model} with parameters $(T, p)$ where $p$ is the map $e 
\mapsto p(e)$.

In keeping with the biological setting we will call an arbitrary partition
$\chi$ of $X$ a {\em character} (on $X$). Let  $\P[\chi|T,p]$ denote the 
probability of generating a character $\chi$
under the random cluster model with parameters $(T,p)$.   We say a subset 
$C$ of the set $E(T)$ of edges of $T$ is a
{\em cutset for $\chi$ on $T$} if the partition $\chi$ of $X$ equals that 
induced by the components of $(V(T), E(T)-C)$.
Then
\begin{equation}
\label{expeq}
\P[\chi|T,p] = \sum_{C}\prod_{e \in C}p(e)\prod_{e \in E(T)-C}(1-p(e)),
\end{equation}
where the summation is over all cutsets $C$ for $\chi$ on $T$.
Note that the number of terms in the
summation described by (\ref{expeq}) can be exponential with $n=|X|$. 
However by modifying the
well-known dynamic programming approach for computing the probability of a 
character
on a tree according to a \underline{finite} state Markov process (see eg. 
\cite{fel81})
one can compute $\P[\chi|T,p]$ in polynomial time in $n=|X|$.

Note that the probability distribution described by (\ref{expeq}) models the evolution of characters under the assumptions that any substitution is always to a new state, and with indepedence between substitution events on different edges of the tree. We will relate this model to more explicit models of character evolution (on a finite but large state space) in the next section.

Suppose we generate a sequence $\Pi= (\chi_1, \ldots,
\chi_k)$ of $k$ such independent 
characters on $X$ where the generating pair $(T,p)$
is unknown. We wish to reconstruct $T$ with probability at least
$1-\epsilon$ from $\Pi$.

The following theorem, from \cite{mos03} describes how the required value of $k$ is related to the
size of $T$ and properties of $p$, and illustrates the logarithm-polynomial phase transition that occurs depending on whether or not the $p(e)$ values are all less than $1/2$.  We refer the reader to \cite{mos03}
for the proof of this result. 

\begin{theorem}
\label{mainthm}
Let $0<l \leq u <1$ and $0<\epsilon<1$ be fixed constants.  Consider the
random cluster model on any collection of the parameters $(T, p)$ where
$T$ is a trivalent phylogenetic tree, and $l \leq p(e) \leq u$ for all
edges $e$ of $T$.  Let $k$ be the number of characters
generated i.i.d. under this model, and $k_{\min}(\eps)$ be the minimal 
$k$ such that the tree can be correctly reconstructed from the
characters with probability at least $1-\epsilon$.
Then, if $n$ denotes the number of leaves of $T$.
\begin{itemize}
\item[(i)] $k_{\min}(\eps)$ grows logarithmically with $n$ if $u <
  \frac{1}{2}$. In particular, if
\[
 k \geq \frac{2(1-u)^4}{l (1-2u)^4}
\log\left(\frac{n}{\sqrt{\epsilon}}\right),
\]
then the tree can be reconstructed correctly with probability 
$1 - \eps$. Furthermore, there is a polynomial-time (in $n$)  
algorithm for reconstructing 
$T$ from the generated characters.

\item[(ii)] $k_{\min}(\eps)$ can grow polynomially with $n$ if $l  >
  \frac{1}{2}$. In particular, for all $h$, if
\begin{equation} \label{eq:low_k}
k \leq \frac{\eps (1-l)^{h}}{6} \left(\frac{n}{3}\right)^{-\log_2(2-2l)},
\end{equation}
then there exists a distribution on trivalent 
phylogenetic $X$-trees, such that if $T$ is drawn according to the
distribution, $p(e) = l$, for all edges of the trees, and 
characters are generated by $(T,p)$, then 
the probability of correctly reconstructing $T$ given the $k$
characters is bounded above by $\eps + 3^{-3 \times 2^h}$.
\end{itemize} 
\end{theorem}

Theorem~\ref{mainthm} shows that the situation with the random cluster
model differs from a bounded-state spaces model such as the CFN model, in two respects. 
Firstly, the critical value
is $\frac{1}{2}$ instead of $\frac{1}{2}(1-\frac{1}{\sqrt{2}})$. 
This corresponds to the fact that in statistical physics models on the
binary tree, 
the critical value for the
extremality of the free measure or the Ising model is
$\frac{1}{2}(1-\frac{1}{\sqrt{2}})$, see \cite{BRZ,EKPS,M:recur}, 
while the critical value for
uniqueness of Gibbs measure, or the critical value for percolation is
$\frac{1}{2}$, see \cite{Grimmett,Pbook}. In \cite{mos2} it is shown that for
any Markov model, if the substitution rate is high then $k$ depends
polynomially on $n$.

The second respect in which the random cluster model differs from the
symmetric two state model, is that for the random cluster model, the
dependence of $k$ on $l$ has exponent $-1$ rather than $-2$, see
\cite{erd99, mos, ste2}.

We now provide a brief outline of the proof of part (i) of
Theorem~\ref{mainthm}, since it combines a combinatorial result with a
probabilistic percolation argument; full details can be found in
\cite{mos03}.  Recall that a {\em quartet tree} is a trivalent
phylogenetic $X$--tree for $|X|=4$. We can represent any quartet tree by
the notation $xy|wz$ where $x,y$ are leaves that are adjacent to one
interior vertex, while $w,z$ are leaves that are adjacent to the other
interior vertex.  For any trivalent phylogenetic $X$--tree, $T$ let
$\Q(T)$ denote the set of quartet trees induced by $T$ by selecting
subsets of $X$ of size $4$.  It is a fundamental result from \cite{col81} that $T$ is
uniquely determined by $\Q(T)$.  Suppose that $T$ is
a trivalent phylogenetic $X$--tree.  We say that $T$ {\em displays} a
quartet tree $xy|wz$ (respectively, a set $\Q$ of quartet trees) if
$xy|wz \in \Q(T)$ (respectively, if $\Q \subseteq \Q(T)$).  For example
the tree $T$ in Fig.~\ref{fig3}(a) displays the quartet tree $12|47$.
For any three distinct vertices $a,b,c$ of $T$ let $\m(a,b,c)$ denote
the {\em median vertex} of the triple $a,b,c$; that is, the unique
vertex of $T$ that is shared by the paths connecting $a$ and $b$, $a$
and $c$ and $b$ and $c$.

A collection $\Q$ of quartet trees is a {\em generous cover} of $T$
if $\Q \subseteq \Q(T)$ and if, for all pairs of interior vertices
$u, v$ there exists a quartet tree $xx'|yy' \in \Q$ for which $u =
\m(x,x',v)$ and $v = \m(u,y,y')$. 
Given a sequence $\C = (\chi_1, \chi_2, \ldots, \chi_k)$ of characters
on $X$,  let
$$\Q(\C) = \{xx'|yy': \exists i \in\{1,\ldots, k\}: \chi_i(x)=\chi_i(x') 
\neq \chi_i(y)=\chi_i(y')\}.$$

One of the main steps in the proof of Theorem~\ref{mainthm} is to
establish the following purely combinatorial result:

\begin{proposition}
\label{dezy}
If $\Q$ is a generous cover of a trivalent phylogenetic $X$--tree $T$ 
then $T$ is the only phylogenetic $X$--tree that displays $\Q$.
\end{proposition}

Using a percolation-style argument, one can then show that provided $k$
is at least as large as that specified in Theorem~\ref{mainthm}(i), an
i.i.d sequence $\C$ of $k$ characters will, with probability at least
$1-\epsilon$ have the property that $\Q(\C)$ is a generous cover of $T$.

A further extension of Proposition~\ref{dezy}, along with a simulation-based study of the
random cluster model has been described recently by \cite{dez}.

\section{Large but finite state spaces}

Finally, we turn to the question of how many characters one needs to
reconstruct a large tree if the characters evolve under a Markov model
on a large but finite state space.  The results of this section are
based on \cite{stepen} where further details can be found.

As mentioned earlier, many processes involving simple reversible models
of change can be modelled by a random walk on a regular graph. To
explain this connection, suppose there are certain `elementary moves'
that can transform each state into some `neighboring' states.  In this
way we can construct a graph from the state space, by placing an edge
between state $\alpha$ and state $\beta$ precisely if it is possible to
go from either state to the other in one elementary move.  The graph so
obtained is said to be {\em regular}, or more specifically {\em
  $d$--regular} if each state is adjacent to the same number $d$ of
neighboring states.

For example, aligned sequences of length $N$ under the $r$--state
Poisson model can be regarded as a random walk on the set of all 
sequences of length $N$ over $R$; here an elementary move involves
changing the state at any one position to some other state (chosen uniformly
at random from the remaining $r-1$ states). 
Thus the associated graph has $r^N$ vertices
and it is $(r-1)N$--regular.  

As another example, consider a simple model of (unsigned) genome
rearrangement where the state space consists of all permutations of
length $N$ (corresponding to the order of genes $1, \ldots, N$) and an
elementary move consists of an inversion of the order of the elements of
the permutation between positions $i$ and $j$, where this pair is chosen
uniformly at random from all such pairs between $\{1, \ldots, N\}$. In
this case the state space has size $N!$ and the graph is $d$--regular
for $d = \binom{N}{2}$.

Both of the graphs we have just described have more structure than mere $d$--regularity. To describe this we recall the concept of a Cayley graph. Suppose we have a (non-abelian or abelian) group $\G$ together with a subset
$S$ of elements, with the properties that $1_\G \not\in S$ and $s \in S \Rightarrow s^{-1} \in S$.  Then 
the {\em Cayley graph} associated with the pair $(\G,S)$ has vertex set $\G$ and an edge connecting 
$g$ and $g'$ whenever there exists some element $s \in S$ for which $g =
g'\cdot s$.  To recover the above graph on aligned sequences of length
$N$ over an $r$--letter alphabet, we may take $\G$ as the (abelian)
group $(\Z_r)^N$ (the product of $N$ copies of the cyclic group of order
$r$) and the set $S$ of all $N$--tuples that are the
identity element of $\Z_r$ except on any one co-ordinate. To recover the
graph described above for unsigned genome rearrangements we may take
$\G$ to be the (nonabelian) group consisting of all $N!$ permutations 
on $N$ letters
 and $S$ to be the elements corresponding to inversions. 
 
 The demonstration that such graphs are Cayley graphs has an important
 consequence - it implies that they also have the following property. A
 graph $G$ is said to be {\em vertex-transitive} if, for any two
 vertices $u$ and $v$ there is an automorphism of $G$ that maps $u$ to
 $v$. Informally, a graph is vertex-transitive if it `looks the same,
 regardless of which vertex one is standing at'. Clearly a (finite)
 vertex-transitive graph must be $d$--regular for some $d$, and it is an
 easy and standard exercise to show that every Cayley graph is
 vertex-transitive (however not every vertex-transitive graph is a
 Cayley graph, and not every regular graph is vertex-transitive).

Suppose that $R$ is a group, and for some subset $S$ (closed under
inverses and not containing the identity element of $R$) we have a rate
matrix $Q$ (as discussed earlier) for which $Q_{\alpha\beta} = q$ if and
only if there exists some element $s \in S$ for which $\beta = \alpha
\cdot s$, otherwise for any distinct pair $\alpha, \beta$ we have
$Q_{\alpha\beta}=0$.  Such a process we will call a {\em group walk
  process (on the generating set $S$)}.  Group walk processes have a
useful property on trees: for each arc $e=(u,v)$ of $T=(V,E)$ consider
the event $\Delta(e)$ that the state that occurs at $v$ is different
from the state that occurs at $u$ (i.e. there has been a net transition
across the edge). Then using the fact that the Cayley graph for $(R,S)$
is vertex transitive, it is easily shown that the events $(\Delta(e), e
\in E)$ are independent.

For these models we then ask how likely it is that a character evolves
without homoplasy on a tree.  This question has been investigated for
the $2$--state Poisson model (and pairs of taxa) by Chang and Kim
\cite{cha}. Here we consider more general processes on a larger state
space, and for many taxa - consequently we obtain bounds rather than the
exact expressions that are possible in the simpler setting of
\cite{cha}. The proof of the following lemma is straightforward (for
details, see \cite{stepen}).

\begin{lemma}
\label{Markovbound2lem}
Let $(X_t; t \geq 0)$ be a group walk process on generating set of size $d$. 
Then, for any two distinct states $\alpha, \beta$, and any values $s,t \geq 0$,
$$\P[X_{t+s} = \beta|X_t= \alpha] \leq \frac{1}{d}.$$ 
\end{lemma}

The following result shows that for a group-walk process if the size of the generating set is much larger than
$2n^2$ (where $n$ is the number of species) then any character generated on a tree with $n$ species will
almost certainly be homoplasy-free on that tree.

\begin{proposition} \label{Markov} 
Suppose characters evolve on a 
phylogenetic tree $T$ according to a group walk process on a generating set of size $d$.
Let $p(T)$ denote the probability that the resulting randomly-generated 
character $\chi$ is homoplasy-free on $T$. 
Then $$p(T) \geq 1- \frac{(2n-3)(n-1)}{d}$$ where $n=|X|$.  
\end{proposition}

\begin{proof}
Consider a general Markov process on $T$ with state space $R$. 
Suppose that for each arc $(u,v)$ of $T$ and each pair
$\alpha, \beta$ of distinct states in $R$, the conditional
probability that state $\beta$ occurs at $v$ given that $\alpha$ occurs at $u$
is at most $p$.  Then, from \cite{sem02} (Proposition 7.1) we have
$p(T) \geq 1- (2n-3)(n-1)p$.  By Lemma~\ref{Markovbound2lem} 
we may take $p = \frac{1}{d}$. The result now follows. 
\end{proof}

We are now ready to state a result for certain Markov processes on large (but finite!) state spaces, which brings together several ideas presented above. Informally, Theorem~\ref{mainthm2} states that for a group walk process, a growth of around $n^2\log(n)$ in the size of the generating set is sufficient (with all else held constant) for producing a sequence of homoplasy-free characters that define $T$.

Let
$p_{\min} = \min\{\P[\Delta(e)]: e \in E\}$, $p_{\max} = \max\{\P[\Delta(e)]: e \in  E\}$, and 
for any $\epsilon>0$ let
\begin{equation}
\label{ceq}
c_\epsilon= \frac{1+ \log(\frac{1}{\sqrt{\epsilon}})}{\beta \epsilon}
\end{equation}
where $\beta =  p_{\min}(\frac{1-2p_{\max}}{1-p_{\max}})^4$.

\begin{theorem}
\label{mainthm2}
Suppose characters evolve i.i.d. on a binary phylogenetic tree $T$ according to a group walk process on a generating set of size $d$, where
$$d \geq c_\epsilon \cdot n^2\log(n)$$
with 
 $c_{\epsilon}$ given by (\ref{ceq}) and with
$p_{\rm max} < \frac{1}{2}$.
Then with probability at least $1-2\epsilon$ we can correctly reconstruct the topology of $T$ by generating an appropriate 
(and logarithmic in $n$) number of characters.
\end{theorem}

\begin{proof}

Let us generate $k= \lceil \frac{2}{\beta}\log(\frac{n}{\sqrt{\epsilon}})\rceil $ characters under a group walk process on a rooted phylogenetic tree.   Consider the event $H$ that all of these characters are homoplasy-free on $T$.  By Proposition~\ref{Markov} we have
$$\P[H] \geq (1- \frac{(2n-3)(n-1)}{d})^k \geq (1-\frac{2}{c_{\epsilon}\log(n)})^{\frac{2}{\beta}\log(\frac{n}{\sqrt{\epsilon}})}.$$
Now, applying the inequality $(1-x)^y \geq 1-xy$ for $x,y>0$, and straightforward algebra gives
$$\P[H] \geq 1- \frac{\epsilon(1+
\log(\frac{1}{\sqrt{\epsilon}}))^{-1}}{\log(n)}[\log(n) + \log(\frac{1}{\sqrt{\epsilon}})] \geq 1- \epsilon, $$
using $\log(n) \geq 1$.

To relate the group walk process to the random cluster model we use a simple coupling argument. First consider
any linear extension of the partial order induced by $T$ on its vertices - i.e. impose a time-scale on the tree.
To each vertex $v$ assign the pair $(\alpha, i)$ where $\alpha$ is the state of the group walk process at $v$,
and $i \in \{0,1,2,...\}$ indicates how often this state has arisen from another state earlier in the tree. Note that this model always generates new states (it is homoplasy-free). Furthermore, let $A(e)$ denote the event that the two vertices in $e$ have different states under the coupled process. Notice that $A(e)$ occurs precisely when 
the two vertices in $e$ have different states under the original process. Thus, the collection
$\{A(e): e \in E\}$ is a collection of independent events.  Consequently
this coupled process is precisely the random cluster model, and we may
identify $\beta$ with the expression $l(\frac{1-2u}{1-u})^4$ in
Theorem~\ref{mainthm}(i). 
 Furthermore the probability that $T$ will be correctly reconstructed
 from $k$ characters produced by the coupled process is at least
 $1-\epsilon$ by Theorem~\ref{mainthm} (recalling that $p_{\rm max} <
 \frac{1}{2}$).

On the other hand the original $k$ characters induce the same partitions as the derived characters whenever event $H$ holds, and $\P[H] \geq 1-\epsilon$. Consequently, by the Bonferroni inequality, 
the joint probability that event $H$ holds and that the $k$ characters 
produced by the coupled process recovers $T$ is at least $1-2\epsilon$. Thus the probability that the original $k$ characters recover $T$ is at least this joint probability, and so at least $1-2\epsilon$, as claimed. 
\end{proof}

We end this section with a remark of caution. Theorem~\ref{mainthm2}
suggests a possible (but flawed) way to avoid the well-known statistical inconsistency of phylogenetic reconstruction methods such as maximum parsimony or maximum compatibility under simple models of site 
substitution.
This approach is based on grouping sites
into $s$--tuples.  For example if the underlying process is the
symmetric $2$--state model, then if we take $s$--tuples of sites we
obtain a group walk process on a generating set of size $2^s-1$. So by
taking $s$ large enough Theorem~\ref{mainthm2} suggests that methods
like compatibility (or maximum parsimony) should be able to correctly
reconstruct a tree with high probability.  The flaw in this argument is
that as $s$ increases, so too does $p_{\rm max}$ and once $p_{\rm max}$
exceeds $\frac{1}{2}$ Theorem ~\ref{mainthm2} is no longer valid. 
In fact one can show that for the $2$--state symmetric model and a tree
with $4$ leaves, the branch length requirements for the statistical
consistency of maximum parsimony on site data is identical to that on
$s$--tuple site data (Theorem 3A of \cite{stepen2}).

\section{Concluding comments}

Stochastic models of character evolution have come to play an central role in phylogenetic genomics. In this chapter we have investigated various aspects of the broad question: how much information do characters tell us about their evolutionary history?  There are two ways to investigate this issue - one approach that is popular in biology and bioinformatics is to use simulation (for example, generating sequences on a tree and then comparing the tree that is reconstructed from these sequences with the original tree) and several authors have taken this approach (eg. \cite{chu92}). Here we have adopted an analytical approach, that gives provable and general bounds on quantities of interest.  This approach has both advantages and disadvantages over a simulation-based one.  Its main disadvantage is that the bounds we obtain may sometimes be far from being `exact', and in that case simulations provide a useful way to detect and quantify this (as was undertaken for the random cluster model in
 \cite{dez}). However analytic approaches have some unique advantages: the results are generic and not specific to  particular parameter settings in the simulation runs; indeed some of the bounds we presented apply to 
\underline{any} tree reconstruction method, something that would be difficult or impossible to investigate using simulations. Furthermore, analytic results can show precisely the functional dependency between quantities of interest (for example the dependence of $k$ on $\log(n)$ in the random cluster model).  Analytic approaches have one more virtue - the arguments used provide insight into how information is preserved and lost, and what methods might be most effective in recovering it. 

Turning to possible future work, we expect that the conjecture(s) outlined in section \ref{logsec} will likely be resolved in the near future. Another challenging mathematical task would be to provide explicit lower bounds on the tree reconstruction probability for maximum likelihood -- the only known bounds (from a more general result in \cite{ste2} that is not specific to phylogeny) involve exponential dependence of $k$ on $n$ and can likely be improved considerably in the phylogenetic setting. For applications, it would also be useful to be able to estimate reconstruction probabilities (or the sequence length required to achieve a given reconstruction probability) from data, perhaps by exploiting bootstrap statistics.  Although there has been some empirical approaches to this problem (eg. \cite{lec94}) little has been rigorously established. 

\subsection{Acknowledgements}  We thank Olivier Gascuel and the referees for helpful comments on an earlier version of this chapter.


\begin{thebibliography}{99}

\bibitem{AF}
Aldous, D. and Fill, J. A.  (2003).
{\it Reversible Markov chains and random walks on graphs}, 
book in preparation. Current version available at 
http://stat-www.berkeley.edu/users/aldous/book.html.


\bibitem{AS}
Alon, N. and Spencer, J. H. (2000).  
{\it The probabilistic method}, second edition. John Wiley and Sons.


\bibitem{AN} 
Athreya, K. B. and Ney, P. E. (1972). 
{\it Branching Processes}, Springer-Verlag.


\bibitem{ban} 
Bandelt, H. J. and Dress, A. W. M. (1986).
Reconstructing the shape of a tree from observed dissimilarity data.  
{\it Advances in Applied Mathematics}, {\bf 7}, 309--343.
  

\bibitem{BRZ}
Bleher, P. M., Ruiz, J. and Zagrebnov, V. A. (1995).
On the Purity of limiting Gibbs state for the Ising model on the Bethe lattice. 
{\it J. Stat. Phys.}, {\bf 79}, 473--482.
  
  
\bibitem{cha} 
Chang J. T. and Kim, J. (1996). 
The measurement of homoplasy: a stochastic view.
In {\it Homoplasy: The recurrence of similarity in evolution}, 
( ed. M.J. Sanderson and L. Hufford), 189--303. 


\bibitem{cha95} 
Charleston, M. and Steel, M. A. (1995).  
Five surprising properties of parsimoniously colored trees. 
{\it Bulletin of Mathematical Biology}, {\bf 57(2)}, 367--375.


\bibitem{chu92}
Churchill, G. A., von Haesler, A. and Navidi, W. C. (1992). 
Sample size for a phylogenetic inference. 
{\it Mol. Biol. Evol.}, {\bf 9}, 753--769. 
  

\bibitem{col81} 
Colonius, H., and Schulze, H. H. (1981).
Tree structures for proximity data. 
{\it British Journal of Mathematical and Statistical Psychology}, {\bf 34}, 167--180.
  
    
\bibitem{CT} 
Cover, T. M. and Thomas, J. A. (1991).
{\it Elements of Information Theory}. John Wiley and Sons.
  
  
\bibitem{dez} 
Dezulian, T. and Steel, M. (2004).
Phylogenetic closure operations and homoplasy-free evolution.
{\it Proceedings of the International Federation of Classification Societies (IFCS)}.
In press. 


\bibitem{dia88}
Diaconis, P. (1988). 
Group representations in probability and statistics. 
{\it Institute of Mathematical Statistics, Lecture Notes-Monograph Series},
{\bf Vol 11}, (ed. Gupta, S. S.). Hayward, California.
 
 
 
 
\bibitem{dur03}
Durrett, R. Shuffling chromosomes. {\em Annals of Applied Probability} (in press). 

\bibitem{erd99} 
Erd{\"o}s, P. L., Sz{\'e}kely, L. A., Steel, M. and Warnow, T. (1999). 
A few logs suffice to build (almost) all trees (I).
{\it Random Structures and Algorithms}, {\bf 14}, 153--184.


\bibitem{EKPS} 
Evans, W., Kenyon, C., Peres, Y. and Schulman, L. J. (2000).
Broadcasting on trees and the Ising Model. 
{\it Annals of Applied Probability}, {\bf 10(2)}, 410--433.

 
\bibitem{eva93} 
Evans, S. N. and Speed, T. P. (1993). 
Invariants of some probability models used in phylogenetic inference. 
{\it Annals of Statistics}, {\bf 21}, 355--377.
    
    
\bibitem{far96} 
Farach, M. and Kannan, S. (1999). 
Efficient algorithms for inverting evolution. 
In {\it Journal of the Association for Computing Machinery}, {\bf 46}, 437--449.
    

    
\bibitem{fel81} 
Felsenstein, J. (1981a).
Evolutionary trees from DNA sequences: A maximum likelihood approach. 
{\it Journal of Molecular Evolution}, {\bf 17}, 368--376.  
    
    
\bibitem{fel04}
Felsenstein, J. (2004). 
{\it Inferring Phylogenies}, Sinauer Press.
  
  
\bibitem{gal02}
Gallut, C. and Barriel, V. (2002).
Cladistic coding of genomic maps.
{\it Cladistics}, {\bf 18}, 526--536.
    
    
\bibitem{gal01} 
Galtier, N. (2001). 
Maximum-likelihood phylogenetic analysis under a covarion-like model. 
{\it Molecular Biology and Evolution}, {\bf 18}, 866--873.
    
    
\bibitem{Georgii}
Georgii, H. O. (1988).
Gibbs measures and phase transitions. 
{\it de Gruyter Studies in Mathematics}, Walter de Gruyter and Co., Berlin, 9.


\bibitem{GHM}
Georgii, H. O., H\"{a}ggström, O. and Maes, C. (2001).
The random geometry of equilibrium phases. 
{\it Phase Transitions and Critical Phenomena}, 
(eds. Domb, C. and Lebowitz, J. L.), Academic Press, London. pp. 1--142.


\bibitem{Grimmett}
Grimmett, G. (1999).
{\it Percolation} (2nd edn). Springer-Verlag, Berlin.


\bibitem{gui} 
Guiasu, S. (1977). 
{\it Information theory with applications}, McGraw-Hill, New York. 


\bibitem{higuchi:77}
Higuchi, Y. (1977).
Remarks on the limiting {G}ibbs states on a {$(d+1)$}-tree.
{\it Publications of the Research Institute for Mathematical Sciences}, {\bf 13(2)}, 335--348.

    
\bibitem{hub}
Huber, K. T., Moulton, V. and Steel, M. (2002).
Four characters suffice to convexly define a phylogenetic tree. 
{\it Research Report UCDMA2002/12}, Department of Mathematics and Statistics, 
University of Canterbury, Christchurch, New Zealand.


\bibitem{JansonMossel:04} 
Janson, S. and Mossel, E. (2003). 
Robust reconstruction on trees is determined by the second eigenvalue. 
to appear in {\it Annals of Probability} 


\bibitem{KestenStigum:66}
Kesten, H. and Stigum, B. P. (1966).
Additional limit theorems for indecomposable multidimensional {G}alton-{W}atson processes.
{\it Annals of Mathematical Statistics}, {\bf 37}, 1463--1481.


\bibitem{kim64}
Kimura, M. and Crow, J. (1964).
The number of alleles that can be maintained in a finite population. 
{\it Genetics}, {\bf 49}, 725--738.


\bibitem{lec94}
Lecointre, G., Philippe, H., Lanh van Le, H. and Le Guyader, H. (1994).
How many nucleotides are required to resolve a phylogenetic problem? 
{\it Molecular phylogenetics and evolution}, {\bf 3}, 292--309.


\bibitem{Martin:03}
Martin, J. (2003).
Reconstruction thresholds on regular trees.
In {\it Discrete Random Walks}, (eds. Banderier, C. and Krattenthaler, C.), 
{\it Discrete Mathematics and Theoretical Computer Science}, 191--204.
\newblock Availible at {\tt http://dmtcs.loria.fr/proceedings/dmACind.html}.


\bibitem{MaSiWe:03b}
Martinelli, F., Sinclair, A. and Weitz, D. (2003).
The ising model on trees: Boundary conditions and mixing time.
Submitted to {\it Communication in Mathematical Physics}. 
Extended abstract appeared in \cite{MaSiWe:03a}.


\bibitem{MaSiWe:03a}
Martinelli, F., Sinclair, A. and Weitz, D. (2003).
The ising model on trees: Boundary conditions and mixing time.
In {\it Proceedings of the Forty Fourth Annual Symposium on Foundations of Computer Science}, 628--639.


 \bibitem{mor1}
Moret, B. M. E., Tang, J., Wang, L.S. and Warnow, T. (2002).
Steps toward accurate reconstruction of phylogenies from gene-order data.  
{\it Journal of Computer and System Sciences}, {\bf 65(3)}, 508--525.


\bibitem{mor2}
Moret, B. M. E., Wang, L. S., Warnow, T. and Wyman, S. (2001).
New approaches for reconstructing phylogenies based on gene order.  
Proc. 9th Int'l Conf. on Intelligent Systems for Molecular Biology ISMB-2001, 
{\it Bioinformatics}, {\bf 17}, S165--S173.



\bibitem{mos} 
Mossel, E. 
Phase transitions in phylogeny. 
{\it Transactions of the American Mathematical Society}, in press.


\bibitem{M:lam2} 
Mossel, E. (2001).
Reconstruction on trees: Beating the second eigenvalue.
{\it Annals of Applied Probability}, {\bf 11}, 285--300.


\bibitem{M:recur} 
Mossel, E. (1998).
Recursive reconstruction on periodic trees.
{\it Random Structures and algorithms}, {\bf 13}, 81--97.
            
            
\bibitem{mos2}
Mossel, E. (2003). 
On the impossibility of reconstructing ancestral data and phylogenies. 
{\it Journal of computational biology}, {\bf 10(5)}, 669--676.

\bibitem{m:forest} E. Mossel, (2004).
Distorted metrics on trees and phylogenetic forests, preprint.
Availible at the Arxiv: http://arxiv.org/abs/math.CO/0403508.


\bibitem{MosselPeres:03}
Mossel, E and Peres, Y. (2003).
Information flow on trees.
{\it Annals of Applied Probability}, {\bf 13(3)}, 817--844.
 
 
\bibitem{mos03}  
Mossel, E. and Steel, M. (2004). 
A phase transition  for a random cluster model on phylogenetic trees. 
{\it Mathematical Biosciences}, 187: 189-203. 


\bibitem{pp}
Pemantale, R. and Peres, Y. (1995).
Recursions on trees and the ising model at critical temperatures.
Unpublished manuscript.
 
 
\bibitem{nad}
Nadeau, J. J. and Taylor, B. A.
Lengths of chromosome segments conserved since divergence of man and mouse. 
{\it Proceedings of the National Academy of Sciences, USA}, {\bf 81}, 814--818.
   
  
\bibitem{pen}
Penny, D., McComish, B.J., Charleston, M. A. and Hendy, M. D. (2001). 
Mathematical elegance with biochemical realism: the covarion model of molecular evolution. 
{\it Journal of Molecular Evolution}, {\bf 53}, 711--723.


\bibitem{Pbook}
Peres, Y.  (1997). 
Probability on trees: an introductory climb.
{\it Lectures on probability theory and statistics (Saint-Flour, 1997)}, 193--280.
Lecture Notes in Math., 1717, Springer, Berlin.


\bibitem{rok}
Rokas, A. and Holland, P. W. H. (2000).
Rare genomic changes as a tool for phylogenetics. 
{\it Trends in Ecology and Evolution}, {\bf 15}, 454--459.

 
\bibitem{sch70} 
Schr{\"o}der, E. (1870). 
Vier combinatorische probleme.
{\it Zeitschrift f{\"u}r Mathematik und Physik}, {\bf 15}, 361--376.

\bibitem{sem02} 
Semple, C. and Steel, M.  (2002).
Tree reconstruction from multi-state characters. 
{\it Advances in Applied Mathematics}, {\bf 28}, 169--184.


\bibitem{sem03} 
Semple, C. and Steel, M. (2003).
{\it Phylogenetics}. Oxford University Press.


\bibitem{sob03} 
Sober, E. and Steel, E. (2002). 
Testing the hypothesis of common ancestry. 
{\it Journal of Theoretical Biology}, {\bf 218}, 395--408. 


\bibitem{ste94} 
Steel, M. (1994). 
Recovering a tree from the leaf colourations it generates under Markov model. 
{\it Applied Mathematical Letters}, {\bf 7}, 19--23.


\bibitem{Stconj}
Steel, M. (2001).
My favourite conjecture, http://www.math.canterbury.ac.nz/$\sim$mathmas/conjecture.pdf.


\bibitem{stepen}
Steel, M. and Penny, D. (2004). 
MP and the phylogenetic information in multi-state characters. (submitted). 

\bibitem{stepen2}
Steel, M. and Penny, D. (2000). 
Parsimony, likelihood and the role of models in molecular phylogenetics.
{\em Mol. Biol. Evol.} 17(6): 839--850. 

\bibitem{sze} 
Steel, M. and Sz{\'e}kely, L. A. (1999).
Inverting random functions (I). 
{\it Annals of Combinatorics}, {\bf 3}, 103--113.

 
\bibitem{ste2}
Steel, M. and Sz{\'e}kely, L. A. (2002).
Inverting random functions (II): explicit bounds for discrete maximum likelihood estimation, with applications. 
{\it SIAM Journal on Discrete Mathematics}, {\bf 15}, 562--575.
  
    
\bibitem{swo96} 
Swofford, D. L.,  Olsen, G. J., Waddell, P. J. and Hillis, D. M. (1996).
Phylogenetic inference. In {\it Molecular Systematics} (2nd edn.), 
(eds. Hillis, D. M, Moritz, C. and Marble, B.K.), Sinauer, Sunderland, U.S.A., 407--514.
    

\bibitem{tuf98}
Tuffley, C. and Steel, M. (1998). 
Modelling the covarion hypothesis of nucleotide substitution. 
{\it Mathematical Biosciences}, {\bf 147}, 63--91.


\end{thebibliography}
\end{document}